\documentclass{article}
\usepackage[utf8]{inputenc}

\usepackage{amssymb,amsmath,amsfonts,amsthm}
\usepackage{pifont}

\usepackage{bm}
\usepackage{lineno}

\usepackage{graphicx}
\usepackage{bbm}
\usepackage{tabularx}
\usepackage{multirow}

\usepackage[super,sort&compress]{natbib}
\usepackage{amsfonts}
\usepackage{booktabs} 
\usepackage{array} 
\usepackage[above, below]{placeins}
\usepackage{multicol}
\usepackage{color}

\usepackage[a4paper,left=1in,top=1in,right=1in,bottom=1in,ignoreheadfoot]{geometry}
\usepackage{algorithm}
\usepackage{afterpage}

\usepackage{framed}
\usepackage{epstopdf}

\usepackage{threeparttable}
\usepackage{pgf}

\usepackage{authblk}

\usepackage{caption}
\usepackage{subcaption}

\usepackage{mathtools}

\DeclarePairedDelimiter\floor{\lfloor}{\rfloor}

\usepackage{natbib}
\usepackage{graphicx}

\begin{document}

\vspace*{0.2in}

\begin{flushleft}
{\Large
\textbf\newline{A decision theoretic approach to model evaluation in computational drug discovery}
}
\newline
\\
Oliver Watson\textsuperscript{1*},
Isidro Cortes-Ciriano\textsuperscript{1,2},
Aimee Taylor \textsuperscript{3,4},
James A Watson\textsuperscript{5,6}
\\
\bigskip
\textbf{1} Evariste Technologies Ltd, Goring on Thames, United Kingdom
\\
\textbf{2} Centre for Molecular Science Informatics, Department of Chemistry, University of Cambridge, Lensfield Road, Cambridge CB2 1EW, United Kingdom
\\
\textbf{3} Center for Communicable Disease Dynamics, Department of Epidemiology, Harvard T.H. Chan School of Public Health, Boston, Massachusetts, United States of America, 
\\
\textbf{4} Infectious Disease Microbiome Program, Broad Institute, Cambridge, Massachusetts, United States of America
\\
\textbf{5} Centre for Tropical Medicine and Global Health, Nuffield Department of Medicine, University of Oxford, United Kingdom
\\
\textbf{6} Mahidol-Oxford Tropical Medicine Research Unit, Faculty of Tropical Medicine, Mahidol University, Thailand
\\
\bigskip

* owatson79@evartech.co.uk

\end{flushleft}


\section*{Abstract}
{\bf Background}
Artificial intelligence, trained via machine learning (e.g. neural nets and random forests) or computational statistics algorithms (e.g. support vector machines and ridge regression), holds much promise for the improvement of small molecule drug discovery.
However, structure-activity data are high dimensional with low signal-to-noise ratios and proper validation of predictive methods is difficult. Controlling over-fitting remains a major concern.
It is poorly understood which, if any, of the currently available machine learning algorithms will best predict new candidate drugs.

{\bf Methods}
25 publicly available molecular datasets were extracted from ChEMBL. Neural nets, random forests, support vector machines (regression) and ridge regression were then fitted to the structure-activity data. A new validation method, based on quantile splits on the activity distribution function, is proposed for the construction of training and testing sets. This is compared to standard random partitioning of the data.
Out-of-sample performance was evaluated using mean squared error and new two rank-based loss functions which penalize only the predicted ranks of high activity molecules.

{\bf Results}
Model validation based on random partitioning of available data favours models which overfit and `memorize' the training set, namely random forests and deep neural nets.
Partitioning based on quantiles of the activity distribution correctly penalizes models which can extrapolate onto structurally different molecules outside of the training data. This approach favours more constrained models, namely ridge regression and support vector regression.
In addition, our new rank-based loss functions give considerably different results from mean squared error highlighting the necessity to define model optimality with respect to the decision task at hand.

{\bf Conclusions}
Model performance should be evaluated from a decision theoretic perspective with subjective loss functions, i.e. loss functions that best encode the modelling goals. Data-splitting based on the separation of high and low activity data provides a robust methodology for determining the best extrapolating model. Simpler, traditional statistical methods such as ridge regression outperform state-of-the-art machine learning methods in this setting.

\section*{Introduction}

Empirical methodologies guide a significant proportion of early stage small-molecule drug discovery \cite{lipinski:04,keiser:07,cumming:13}. These range from simple rule based methods (Lipinski's rule of 5), to searching over molecules `similar' to those already known, to using more complex regression models.
This work concerns the objective evaluation of the predictive ability of the latter, namely statistical and machine learning regression models trained on molecular structure-activity data.
The goal of these models is to characterize the relationship between a high-dimensional binary vector representation of small molecules (molecular fingerprint) and their corresponding target specific \textit{in vitro} activities. In this context, use of regression modeling is often known as quantitative structure-activity relationship modeling (QSAR) \cite{sliwoski:14,van:03}, and many different model classes have been used: support vector machines \cite{burbidge:01}, ridge regression \cite{nandi:07}, neural nets \cite{ajay:98,sadowski:98,lenselink:17,nandi:07} and random forests \cite{svetnik:03}, to name but a few. The success of these models is in part due to high-throughput screening experiments which produce large structure-activity datasets (order of magnitude $10^2$-$10^6$ data-points).

Regression with high-dimensional bioinformatic data is known to be difficult. 
Problems include \textit{the curse of dimensionality}, optimization bias, reporting bias, and low signal-to-noise ratios \cite{Jelizarow:10,Boulesteix:09,ioannidis:05,somorjai:03,Zervakis:09,castaldi:11,matveeva:16}. The main theoretical framework which underpins the use and interpretation of these methods is cross-validation \cite{stone:74,geisser:75}, which provides an estimate of the predictive error rate \cite{castaldi:11,molinaro:05}.
However, validation strategies based on random partitioning of datasets, either by $K$-fold cross-validation or the bootstrap, are known to be optimistic for structure-activity modelling \cite{sheridan:13, wu:18,wallach:18}. Multiple alternative strategies have been proposed, for example, splitting by date of assay \cite{sheridan:13}, constructing local neighbourhoods based on similarity scores or scaffold splitting \cite{sheridan:13,wu:18}, or stratified sampling whereby equal distributions of the activity levels are assured across training and testing sets \cite{wu:18}.

This work also argues that standard validation approaches - $K$-fold cross-validation and the bootstrap - which based on random partitioning of available data, will not target the true predictive model error in the context of small molecule drug discovery. We give a theoretical justification for this claim and show it empirically using 25 publicly available datasets. We propose a simple alternative partitioning method which splits datasets on quantiles of the activity distribution function. This univariate parametrisation of the  training set construction allows for inference on the predictive ability of different regression methods in the limit: as information in the training set is reduced to almost zero.
In addition, we argue that out-of-sample model performance should be evaluated from a decision theoretic perspective \cite{savage:54} using loss functions which reflect as best possible the process of drug discovery.
Tailor-made loss functions will better determine truly optimal model classes compared to standard goodness-of-fit metrics. We propose simple rank based loss functions to evaluate out-of-sample model prediction accuracy. We show that in these low signal-to-noise settings \cite{cortes:16,kramer:12,kalliokoski:13,kalliokoski:13b}, models with greater structural constraints (ridge regression and linear kernel support vector regression) outperform less constrained machine learning algorithms (neural nets and random forests). 

\section*{Methods}

\subsection*{Cross-validation with biased data}

\subsubsection*{Problem setting}

This section outlines the formal framework and notation we use throughout the paper. We consider the general problem of comparing the performance of multiple predictive models (statistical or machine learning) with respect to a given dataset.
`Optimality' of these predictive models is evaluated with respect to a well defined loss function.

The context investigated here is finding `active' molecules within molecular space. `Active' is defined as having activity level above a given threshold. This activity is target specific. Conditional on a given initial dataset, the overall loss (negative utility of the model) is defined as a function of the number of new molecules needed to be tested until an active molecule is reached.

Each molecule is represented by its `molecular fingerprint', a $P$-dimensional binary vector. We denote this as ${\bf x_i} = \{x_i^j\}_{j=1}^P$, where $i$ indexes the molecule and $j$ indexes the feature (as referred to in the machine learning literature) or covariate (statistics literature).
We have $P=128$ for the fingerprint representation used in this analysis.
Each molecule ${\bf x_i}$ has a target specific activity $y_i$ which corresponds to the negative logarithmic \textit{in vitro} half-inhibitory concentration (p-IC\textsubscript{50}: higher values correspond to increased activity).
In this section we ignore the target specificity as each dataset has an associated target and the datasets are analyzed independently. We do not consider multi-objective regression models here.
We denote the (unknown) functional relationship between the fingerprint and the outcome (activity) as $y = G({\bf x}) + \epsilon$, where $\epsilon$ is experimental error. 

Given a choice of models $M_1,..,M_T$, respective performances are commonly evaluated using $K$-fold cross-validation \cite{stone:74,molinaro:05} (detailed description given in \cite{friedman:01}, Chapter 7), or the bootstrap \cite{efron:83}, closely related strategies.
Standard $K$-fold cross-validation proceeds by dividing the data $\{({\bf x_i},y_i)\}_{i=1}^N$ into a partition of $K>1$ equally sized subsets $S_1,..,S_K$. For the $k^{th}$ subset, we train (fit) our model $M_t$ using the data $S_{train} = \bigcup_{m \neq k} S_m$. The out-of-sample expected loss is then estimated by $l_k = L\left[ \{y_i\}_{i \in S_k}, M_t(\{y_i\}_{i \in S_k} | S_{train})\right]$.
The overall expected loss estimate is $\frac{1}{K} \sum_{k=1}^K l_k $. The notation for the expected loss over each test set is deliberately not summed over the indices of the testing data as this paper considers non-additive loss functions, e.g. aggregate functions of the testing data. The choice of the number of folds $K$ is context dependent and relates to a bias-variance trade-off: smaller $K$ implies a smaller training set and thus increased positive bias in the error rate estimate, however smaller $K$ also forces greater dis-similarity between the training sets and thus lowers variance in the overall error estimate.
The bootstrap is similar to 3-fold cross-validation, whereby two-thirds of the data are used in the training set taken as a bootstrap sample of size $N$ (sampling with replacement). Predictive error estimation is then done by averaging the out-of-bag errors. The bootstrap generally improves on standard $K$-fold cross-validation as it smooths the predictive error when using discontinuous loss functions.

$K$-fold cross-validation and the bootstrap provide nearly unbiased estimators of the conditional expected loss if the empirical distribution $\Hat{F_x}$ (in this context ${\bf x}$ denotes a molecule) is an i.i.d. draw from the true underlying data generating distribution \cite{devroye:96}.
In applications where the goal is to accurately predict the outcome of new data drawn at random with respect to a given data generating process, these are the correct methods for selecting an optimal predictive model. However, drug (lead candidate) discovery is better thought of as a complex optimization problem rather than a passive data prediction problem. The goal here is to generalize (extrapolate) from a model trained on a relatively small dataset to find active molecules in a high dimensional space ($2^P$ possibilities in total). 

If the data-generating distribution (e.g. the distribution that has provided the data at hand: think of this as the protocols which have lead to the data-generating assays) is substantially different from the uniform distribution over the subset of feasible molecules within the $2^P$ possibilities (this is an unknown subset), then these validation methods can give a biased estimate of the true out-of-sample loss \cite{braga:14,wood:07}. For example, the data might be clustered together (with respect to Manhattan distance over the space of fingerprints) and therefore the out-of-sample estimate may in fact be highly skewed towards the in-sample estimate, leading to overconfidence. Molecular structure-activity datasets are highly likely to be skewed in this way \citep{wallach:18} and therefore it is necessary to partition the data in such a way that the out-of-sample testing subset is truly distinct from the in-sample data, thus giving reliable expected loss estimates which do not favour models that over-fit to the training data. This partition should also reflect the decision problem at hand. 

We next describe non-random data partitions which corrects for clustering of training data.

\subsubsection*{Activity dependent model validation}

The subset of feasible molecules is unknown and this makes it difficult to determine whether a given training set is `close' to a test set. `Close' needs to be defined with respect to a particular metric.
Metrics such as the Manhattan distance may be a poor proxy of this true (target specific) distance between subsets of data.
Instead, we propose using the observed outcome (activity) $y$ as the discriminant measure between molecules. Data partitions based on the activity function $G$ (function relating the molecular fingerprint to the p-IC\textsubscript{50}) instead of random partitions force dis-similarities between subsets in the partition. If $G({\bf x_1}) >> G({\bf x_2})$, we assume that ${\bf x_1}$ is notably experimentally different from ${\bf x_2}$.

The following validation design is proposed. Let $\Hat{F_y}$ be the empirical distribution over the activities $\{y_i\}_{i=1}^N$. Let $q\in (0,1)$ be a fixed fraction of the data used to determine the training set. 
With respect to the empirical distribution $\Hat{F_y}$, this maps onto an activity threshold $Y_{q}$ (the q\textsuperscript{th} quantile of $\Hat{F_y}$). The training set is then constructed by bootstrapping the molecules with activity less than $Y_q$. This is the opposite of standard balanced or stratified cross-validation where one assures equal distributions of outcomes across the testing folds \cite{breiman:84}. This is not a `cross-validation' design as the test data are never used as training data.

Multiple bootstrapped iterations are then computed in order to construct confidence intervals around the out-of-sample expected loss estimate \cite{efron:83}. This can be thought of as a stabilizing process within the validation procedure.

In the following, we assume that the molecule index corresponds to the rank of the activities: $y_1 \leq y_2 \leq.. \leq y_N$. Let $N_q = \floor{N\times q}$ be the number of elements in the training set based on the q\textsuperscript{th} quantile.

For each model $M_t$, evaluate for $a = 1,..,A$ independent iterations:
\begin{itemize}
    \item Sample with replacement $N_q$ elements from $\{{\bf x_i} \}_{i=1}^{N_q}$ to get a bootstrapped training dataset $X^{q}_a$. 
    \item Compute $l_a = L\left[ \{y_i\}_{i = N_q +1}^N, M_t \left(\{y_i\}_{i=N_q + 1}^{N}| X^{q}_a\right)\right]$, where two proposals for the loss function $L$ are given in the next section.
\end{itemize}

The set $\{l_1,..,l_A\}$ is then used to estimate the mean expected loss and the 95\% confidence intervals.

\subsubsection*{`Active-rank' loss function}

In the context of using statistical or machine learning methods for novel compound drug discovery, out-of-sample performance should not directly map onto standard goodness-of-fit measures (e.g. $R^2$ or mean squared error), but has a simpler decision theoretic interpretation. If these models are to be used in a real setting then a prediction of high activity for a given feature vector (fingerprint) would lead to a physical experiment confirming or refuting this prediction. As stated above, the goal is to find molecules with an activity above a certain threshold (this will be target specific) and therefore each bad prediction (whereby the true activity is less than the threshold) corresponds to incurring a fixed loss (opportunity-cost and cost of experiment). In reality, experimental costs will not be constant (some molecules are more expensive to make than others), however, we simplify the situation to one where each experiment is considered to have a fixed cost. 
In the out-of-sample predictions, minimizing the loss corresponds ranking the active molecules highest. When evaluating the performance of multiple models fitted to a given dataset, if there is one active molecule and a large number of inactive molecules, the expected loss is insensitive to the ranking of all the inactives below the rank of the active(s). The model's fine-grained accuracy in the region of the inactives is of no importance. This is in contrast to standard measures of predictive accuracy and loss such as $R^2$, mean squared error, or receiver operating characteristics (AUC) which have previously been used in this context \cite{giguere:13,svetnik:03,cumming:13,wu:18,sheridan:13,wallach:18}.

We define our `active-rank' loss function as follows. We choose a fraction $\gamma\in(0,1)$, corresponding to a threshold activity $Y_{\gamma}$ with respect to the empirical distribution function $\hat{F_Y}$. In practice $\gamma$ would be close to 1 (e.g. in the range 0.9-0.99) to simulate scenarios where actives molecules are rare and inactives common. The subset of molecules $\{ {\bf{x_i} }\}_{i=N_{\gamma}}^N$ are then defined as `actives'.
We define $N_{\gamma} = \floor{N\gamma}$ (the total number of actives), and $N_{test} = N - N_q$ (the size of the test set).

For the model $M_t$ fit to the training data $X^q$, the out-of-sample loss is defined with respect to the ranks assigned to the out-of-sample active molecules. We take as convention that the ranks assigned to the test data go from 0 (molecule with highest predicted activity) to $N_{test}-1$ (molecules with least predicted activity). The loss which only depends on the rank of the highest ranked active is defined as:

\begin{eqnarray}
\label{eq:Loss1}
L_{\min}^{\gamma}
= \frac{1}{N_{test} - N_{\gamma}} \min_{j = N_{\gamma} .. N_{test}} \mathrm{Rank}_{M_t}( {\bf x_j} )  
\end{eqnarray}

The minimum active rank will vary from 0 (an active molecule is ranked top in the test data), to $N_{test} - N_{\gamma}$ (all the $N_{\gamma}$ active molecules are ranked last). We normalize to obtain a loss function defined over the interval $[0,1]$.
An alternative version of this loss, whereby all the ranks of the active molecules are taken into account, thus penalizing sub-optimal ranking for all active molecules, is given by:

\begin{eqnarray}
\label{eq:Loss2}
L_{\mathrm{sum}}^{\gamma}
=\frac{1}{N_{\gamma} (N_{test}- N_{\gamma}) } \left(\sum_{j = N_{\gamma}}^{N_{test}} \mathrm{Rank}_{M_t}( {\bf x_j} ) - N_{\gamma} (N_{\gamma}-1)/2 \right) 
\end{eqnarray}

The sum of the active ranks will vary from $N_{\gamma} (N_{\gamma}-1)/2$ (all actives are ranked in the top $N_{\gamma}$ molecules) to $N_{\gamma} (N_{test}- N_{\gamma})$ (all actives are ranked in the last $N_{\gamma}$ molecules).

We note that when $N_{\gamma} = 1$, e.g there is only 1 active molecule, $L_{\min}^{\gamma} = L_{\mathrm{sum}}^{\gamma}$.

As mentioned above, both these loss functions are non-additive with respect to the testing data.

\subsection*{Statistical analysis}

All statistical analyses were done in Python version 2.7. 
The entire analysis is fully reproducible via a publicly available Python Jupyter notebook found at https://github.com/owatson/QuantileBootstrap.

\subsubsection*{Regression models}

We evaluated the performance of four model classes: 
\begin{itemize}
\item Support vector regression (Python module: \textit{sklearn}, function \textit{SVR})
\item Random forests (Python module: \textit{sklearn}, function \textit{RandomForestRegressor})
\item Linear ridge regression (Python module: sklearn, function \textit{Ridge})
\item Deep neural networks (Python module: sklearn, functions \textit{Pipline}  and \textit{StandardScalar},  and Python module: keras, function \textit{KerasRegressor}) 
\end{itemize}
 
For support vector regression we used a linear kernel.
For random forests we used the default parameter settings, growing 100 trees each with a maximum tree depth of 10 splits. 
For linear ridge regression we used a penalty term of $\alpha=0.1$.
For deep neural networks we first standardised the data, then used two dense layers, the first of dimension 128 (to match the input feature dimension) and then dimension 16, both with relu activation.

These correspond to standard default choices in the literature. These four model classes are all somewhat insensitive to tuning parameters. 
In order to minimize any optimization bias, we did not attempt to tune any of the parameters to the set of datasets at hand, except for the deep learning models, where we first fit 4 model structures and chose the model with lowest out-of-sample mean squared error using bootstrapped data.


\subsubsection*{Model comparison}

We first compared model performances using 5-fold cross-validation (this uses 80\% of data chosen at random to predict the remaining 20\%) and bootstrapping (this uses approximately two thirds of the data to predict the remaining third). With discontinuous loss functions, bootstrapping smooths the out-of-sample error predictions \cite{efron:83}.
The out-of-sample predictions were evaluated using mean squared error, and both active-rank loss functions $L_{\min}^{\gamma}, L_{\mathrm{sum}}^{\gamma}$. 
For the active-rank loss functions, we evaluated out-of-sample loss using 3 separate $\gamma$ thresholds corresponding respectively to labelling 10, 5 and 1\% of the test data as active.

We then ran our activity dependent validation procedure using progressively lower fractional thresholds for the training data: $q=0.9;0.8;0.6;0.4$. The same three $\gamma$ thresholds were used to evaluate the out of sample expected losses for the active-rank loss functions. All predictions were evaluated with mean squared error and both active-rank loss functions.

Overall performance was evaluated by assuming independence between the 25 datasets.
The total model score assigned to each model $M_t$ is defined as the sum over all datasets of the probabilities that the $M_t$ had lowest expected loss (probability of optimality).
As the number of bootstrap iterations is much lower than then total number of possible iterations ($N_q^{N_q}$), we use the jackknife to calculate the standard error on the mean out-of-sample prediction. 400 bootstrap iterations were used for each model and set of problem definition parameters, i.e. the pair of parameters $(q,\gamma)$.

\subsection*{Data}

\subsubsection*{Data curation}
We extracted IC\textsubscript{50} data from ChEMBL database version 23 for 25 diverse protein targets and receptors.
In order to assemble high-quality data sets,
we only considered IC\textsubscript{50} values for compounds that satisfied the following filtering criteria: (i) an activity unit equal to `nM', (ii) activity relationship equal to `=', (iii) target type equal to SINGLE PROTEIN, and (iv) organism equal to {\it Homo sapiens}. 
Bioactivity values were modeled in a logarithmic scale ({\it i.e.}, pIC\textsubscript{50} $= \log_{10} \mathrm{IC}_{50}$). The average pIC\textsubscript{50} value was calculated for protein-compound pairs with multiple IC\textsubscript{50} measurements available.  

Further details about the data sets are provided in Table \ref{dataTable}. 
A comparative analysis of these datasets has previously been done in the context of iterative model fitting \cite{cortes:18}. 
All data used in this paper (activity levels, 128-bit fingerprints and smiles) are available at:

https://github.com/owatson/QuantileBootstrap.

\begin{table}[]
\centering
\caption{25 publicly available datasets extracted from ChEMBL and analysed in this paper.}
\label{dataTable}
\resizebox{\textwidth}{!}{%
\begin{tabular}{l l c c c }
Target preferred name                    & Target abbreviation  & Uniprot ID & ChEMBL ID  &  \#Bioactive molecules       \\
\hline
Alpha-2a adrenergic receptor             & A2a                  & P08913     & 1867 & 203             \\
Tyrosine-protein kinase ABL              & ABL1                 & P00519     & 1862 & 773  \\
Acetylcholinesterase                     & Acetylcholin         & P22303     & 220  & 3159 \\
Androgen Receptor                        & Androgen             & P10275     & 1871 & 1290 \\
Serine/threonine-protein kinase Aurora-A & Aurora-A             & O14965     & 4722 & 2125 \\
Serine/threonine-protein kinase B-raf    & B-raf                & P15056     & 5145 & 1730 \\
Cannabinoid CB1 receptor                 & Cannabinoid          & P21554     & 218  & 1116 \\
Carbonic anhydrase II                    & Carbonic             & P00918     & 205  & 603  \\
Caspase-3                                & Caspase              & P42574     & 2334 & 1606 \\
Thrombin                                 & Coagulation          & P00734     & 204  & 1700 \\
Cyclooxygenase-1                         & COX-1                & P23219     & 221  & 1343 \\
Cyclooxygenase-2                         & COX-2                & P35354     & 230  & 2855\\
Dihydrofolate reductase                  & Dihydrofolate        & P00374     & 202  & 584 \\
Dopamine D2 receptor                     & Dopamine             & P14416     & 217  & 479  \\
Norepinephrine transporter               & Ephrin               & P23975     & 222  & 1740 \\
Epidermal growth factor receptor erbB1   & erbB1                & P00533     & 203  & 4 868 \\
Estrogen receptor alpha                  & Estrogen             & P03372     & 206  & 1705 \\
Glucocorticoid receptor                  & Glucocorticoid       & P04150     & 2034 & 1447 \\
Glycogen synthase kinase-3 beta          & Glycogen             & P49841     & 262  & 1757 \\
HERG                                     & HERG                 & Q12809     & 240  & 5207 \\
Tyrosine-protein kinase JAK2             & JAK2                 & O60674     & 2971 & 2655 \\
Tyrosine-protein kinase LCK              & LCK                  & P06239     & 258  & 1352\\
Monoamine oxidase A                      & Monoamine            & P21397     & 1951 & 1379 \\
Mu opioid receptor                       & Opioid               & P35372     & 233  & 840  \\
Vanilloid receptor                       & Vanilloid            & Q8NER1     & 4794 & 1923
\end{tabular}%
}
\end{table}

\subsubsection*{Molecular representation}
The python module \textit{Standardiser} was used to standardise all chemical structures. 
Inorganic molecules were removed, and the largest fragment was kept in order to filter out counterions.

We computed circular Morgan fingerprints 52 using RDkit (release version 2013.03.02).
The radius was set to 2 and the fingerprint length to 128.

\section*{Results}

\subsection*{Model performance evaluated using random data partitioning}

\begin{figure}
\centering
\includegraphics[scale=.34]{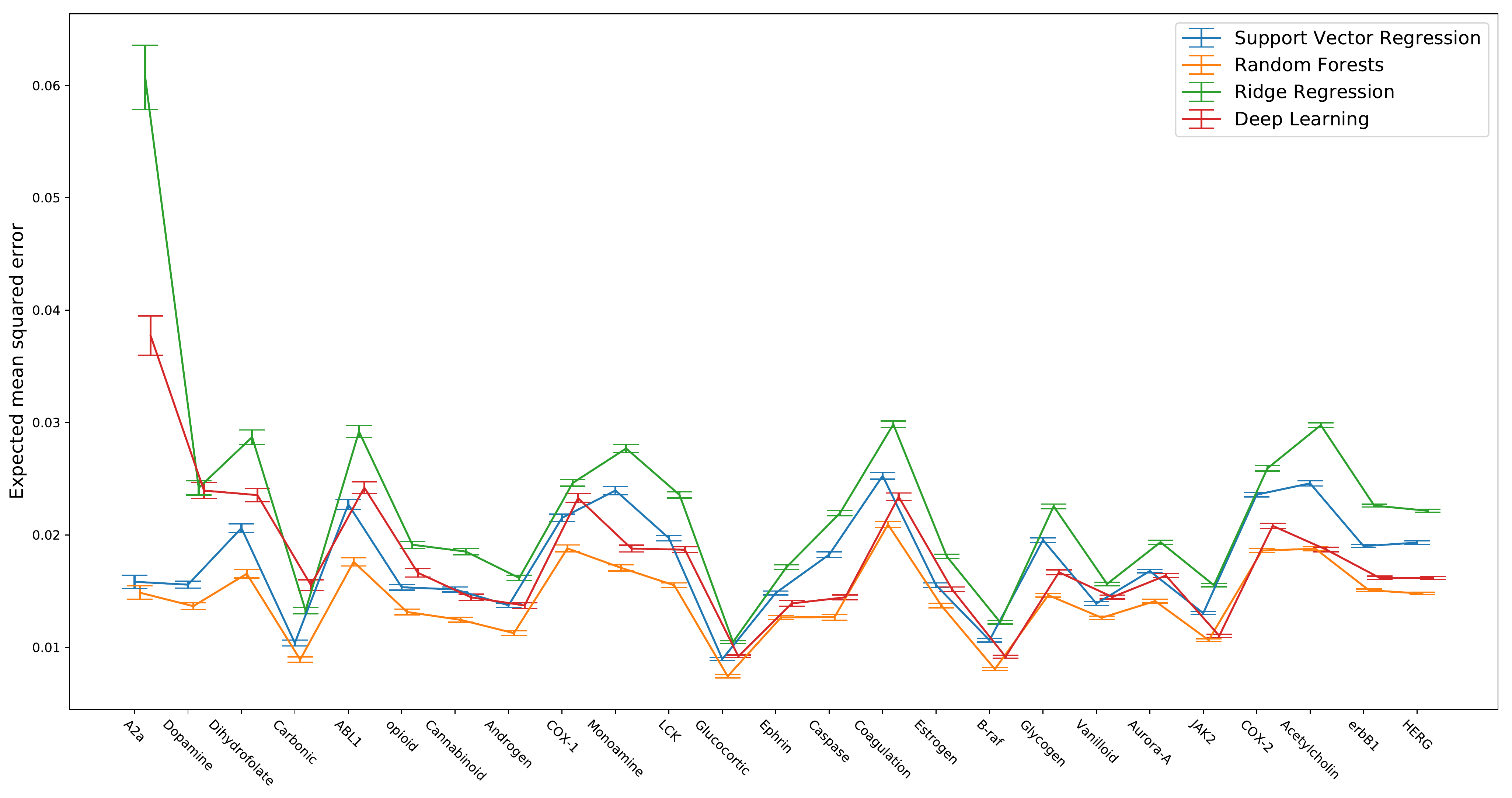}
\caption{{\bf Model comparison using the standard bootstrap.} Expected model out-of-sample mean squared error shown for each dataset, ordered from left to right by increasing size of dataset. Error bars correspond to $\pm$ 2 standard errors around the expected loss estimate, computed using the jackknife estimator. For each datatset, the optimal model is the one with least expected loss, with random forests scoring best for every single dataset.}
\label{Example_LossMSE_Boostrap}
\end{figure}

\begin{figure}[!h]
\centering
\includegraphics[scale=.75]{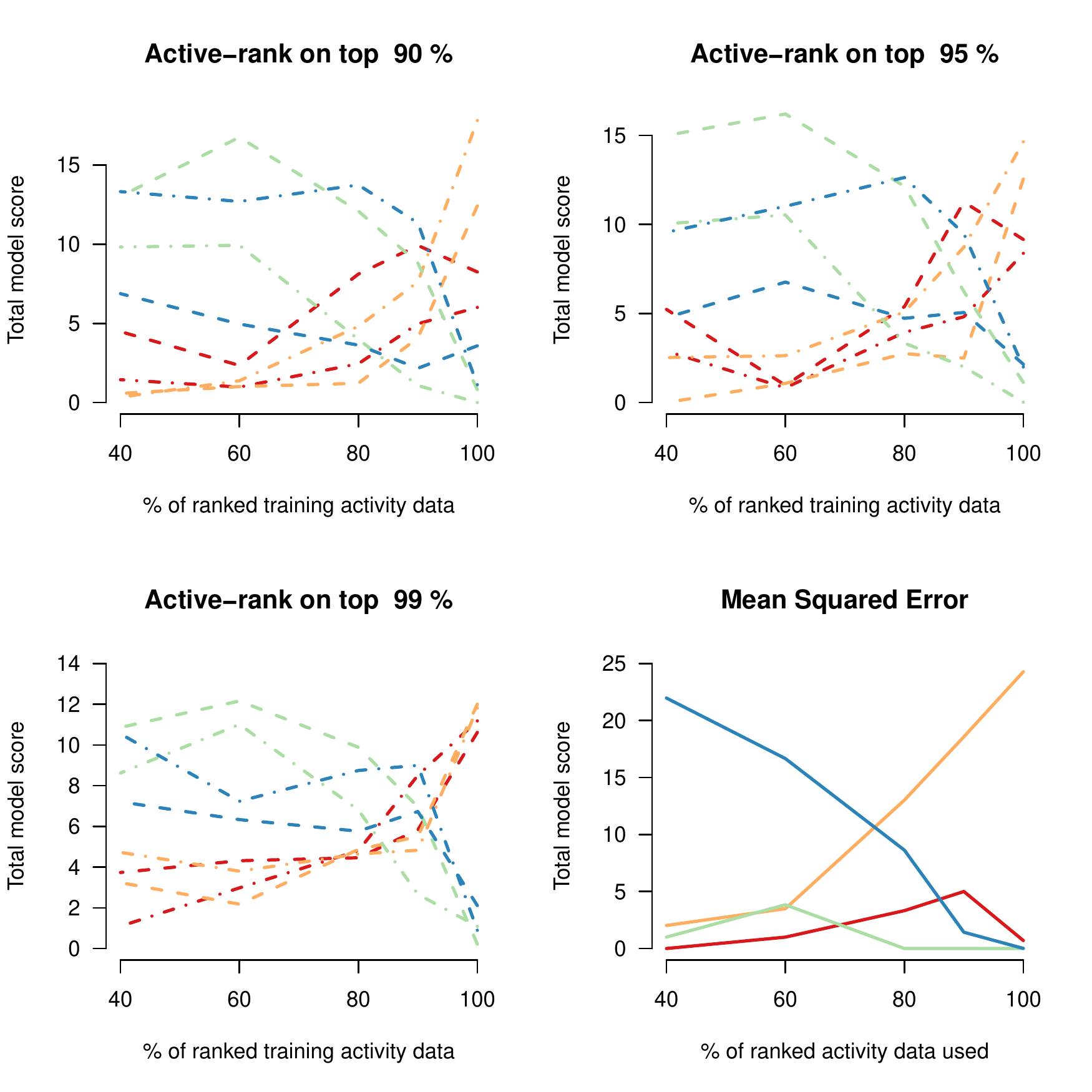}
\caption{{\bf Comparison of overall model performance for the standard bootstrap and the restricted activity bootstrap.} All four panels show the overall model score (sum of the probabilities of model optimality over the 25 datasets) as a function of the restriction on the activity levels in the training data. 100\% corresponds to standard cross-validation (random partitioning). The first three panels show the results for the active-rank loss functions ($L_{\min}^{\gamma}$ shown by dotted lines; $L_{sum}^{\gamma}$ shown by dot-dash lines) with values of $\gamma$ going from 0.9 (top left) to 0.99 (bottom left). The bottom right panel shows the results when models are scored using mean squared error. Red: deep learning; blue: support vector regression; orange: random forests; green: ridge regression.}
\label{overallResults}
\end{figure}

Random partitioning of the data, either using 5-fold cross-validation (training set contains 80\% of the data) or bootstrapping (training set contains two thirds of the data) clearly shows that random forests and deep learning have the best out-of-sample performance (for full results see github Jupyter noteboook). The bootstrap partitioning has the advantage of variance reduction showing clearer trends over the 25 datasets (Figure \ref{Example_LossMSE_Boostrap}).
Random forests performs best on almost every dataset when scored using mean squared error (Figure \ref{overallResults}, bottom right panel), and performs on average as well as deep learning when scored with the active-rank loss functions (Figure \ref{overallResults}, first 3 panels). 
Ridge regression and support vector regression are never optimal in this setting across the 25 datasets, irrespective of the loss function. Figure \ref{Example_LossMSE_Boostrap} shows the bootstrap out-of-bag performance as evaluated by mean squared error for the 4 models over the 25 datasets, with datasets ordered from smallest to largest. Ridge regression has largest out-of-bag error, followed by support vector regression and then deep learning and random forests. This ranking holds for every dataset.

These out-of-sample performances closely reflect the in-sample error. Both deep learning and random forests can almost `memorize' the data with in-sample losses close to zero (see Jupyter notebook).

\subsection*{Target activity dependent cross validation}

Decreasing the quantile activity threshold for the training data from 1 (all data are used in the bootstrap construction of the training set) to 0.4 (only 40\% of the data ordered by activity are used in the bootstrap construction of the training set) results in a complete reversal of optimality amongst the 4 predictive models. When scoring models by out-of-sample mean squared error, support vector regression becomes optimal for quantiles below 0.75 (Figure \ref{overallResults}, bottom right panel).

\begin{figure}
\centering
\includegraphics[scale=.34]{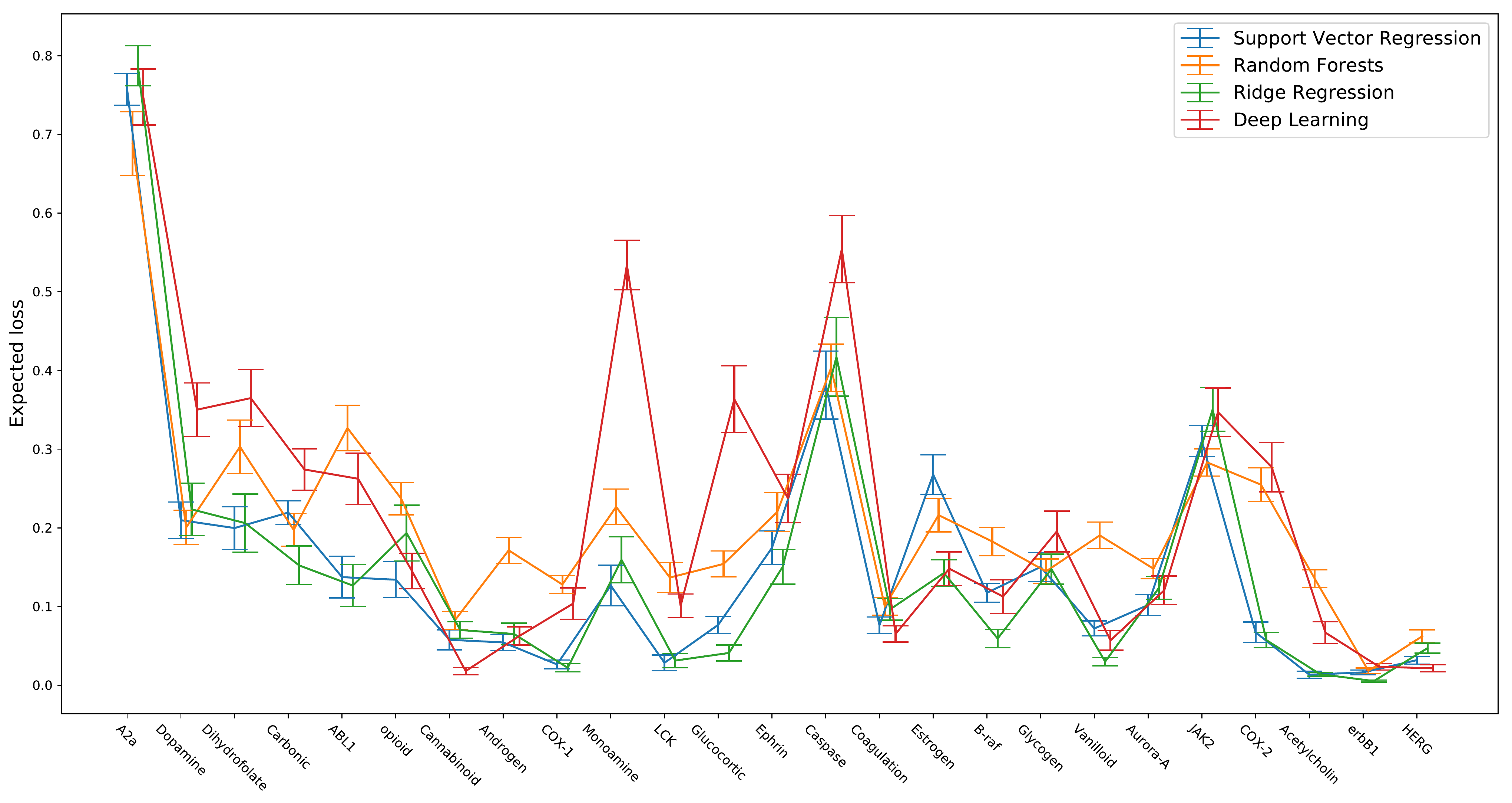}
\caption{{\bf Model comparison using the restricted activity bootstrap with $\gamma=0.4$.} Model expected out-of-sample $L_{min}^{\gamma=0.99}$ loss shown for each dataset, ordered from left to right by increasing size of dataset. Error bars correspond to $\pm$ 2 standard errors around the expected loss estimate, computed using the jackknife estimator. For each datatset, the optimal model is the one with least expected loss.}
\label{Example_Lossmin99}
\end{figure}

For the active-rank loss functions, lowering the activity training threshold also induces a reversal of model optimality (change-point for $q\approx 0.8$). In the most extreme setting ($q=0.4$), support vector regression and ridge regression perform approximately equally well, with total scores corresponding to optimality on half of the datasets. This performance is shown in detail in Figure \ref{Example_Lossmin99}, with datasets ordered from left to right by increasing size. There is some heterogeneity between the datasets for model optimality, but the overall trends are clearly in favour of ridge regression and support vector regression.

By averaging over the 25 datasets, we can see that these trends are robust with respective the target used as the outcome measure in the regression models.

\subsection*{Importance of the loss function}

There are clear disparities between model evaluations for the different loss functions. Mean squared error favours random forests in the standard setting ($q=1$), and support vector regression in the restricted activity setting ($q<0.6$).
However, the active-rank loss functions favour equally deep-learning and random forests in the standard setting, and support vector regression and ridge regression in the restricted activity setting. Moreover, the results differ between $L_{\min}^{\gamma}$ and $L_{sum}^{\gamma}$. The out-of-sample performance of ridge regression is consistently better when evaluated using $L_{\min}^{\gamma}$, and that of support vector regression is consistently better when evaluated using $L_{sum}^{\gamma}$  (Figure \ref{overallResults}).
These results show that the evaluation of model performance is highly dependent on the loss function used. This directly reflects how the different loss functions penalize predictive performance, with $L_{\min}^{\gamma}$ only penalizing the rank of the first active molecule.

\section*{Discussion}


There is considerable hype around the use of AI to find novel drug candidates and to optimize early-stage drug discovery \cite{fleming:18}. Deep-learning via the use of deep neural networks is a highly active research area with a wide range of applications and proven success stories. However, neural networks are known to be extremely `data-hungry' and work best in high signal-to-noise settings \cite{marcus:18}. For regression modelling using molecular structure-activity data, we do not believe deep-learning models will perform well in predicting novel areas of molecular space of high activity, contrary to recent claims \cite{lenselink:17}. 
This modelling exercise shows that partitioning on quantiles of the activity distribution, and thereby mimicking the process of extrapolating onto previously unseen areas of molecular space, removes all predictive power from the deep learning models.
This approach can be contrasted with `temporal splitting' whereby datasets are partitioned by assay date, the first section used to train the model, the second to test.
Although this makes intuitive sense and could be argued to mimic real-life settings, it doesn't guarantee that highly similar molecules - both in structure and activity - will not be found across both testing and training data. Nor does it directly test the capability of a statistical or machine learning model to detect signal predicting activity gradients, resulting in good predictions of molecules with high activity. Splitting on activity quantiles deals with these issues, and provides a simple and interpretable univariate parametrisation of the information content used to train the model.

Evaluation of the predictive performance of regression models when applied to small molecule structure-activity datasets necessitates different approaches than in the standard bioinformatic and high-dimensional settings. Online prediction problems (e.g. image classification, spam filtering, recommender systems, etc) and statistical inference problems (e.g. genome-wide studies, biomarker discovery, micro-array analysis) have different goals. In the drug discovery context, we start with a small training set ($N << 2^P$) and attempt to extrapolate outside of these data in order to find molecules which are inherently `different' from those in the training data. In the machine learning and computational statistics literature, this is most similar to an optimization problem or gradient ascent problem.
This search procedure is done in a relatively resource constrained setting (cost of experimentation, time cost) and therefore model evaluation should be decision theoretic with a subjective loss \cite{savage:54}.

We expect our active-rank loss functions to differ in performance from standard machine learning type losses (most commonly this would be mean squared error). The active-rank loss functions $L_{\min}^{\gamma}$ and $L_{\mathrm{sum}}^{\gamma}$ do not penalize bad predictions outside of the subspace of interest, i.e. high activity areas of molecular space. However, a current limitation of the proposed methods is that the loss functions are non-additive and therefore cannot be used to penalize model fitting in the training phase. This is subject of future work.
Other limitations of the work are that we have done little to no internal model parameter tuning, except for deep neural nets to assess structures most appropriate for these type of data. 
However, we do not expect parameter tuning to considerably change the results nor the conclusions of the study. 
Furthermore, all the analyses are easily reproducible with our openly available Jupyter notebook, thus easily extended to new computational algorithms, different parameter settings, or new datasets.
Lastly, the loss functions used to evaluate model performance on these benchmark datasets will not estimate the true out-of-sample expected loss in experimental settings. In reality, true $\gamma$ thresholds (percentage of feasible molecules above a certain activity level) could be multiple orders of magnitude larger than those used in our study (e.g.  the top $10^{-10}$\% of the  testing data).

\section*{Acknowledgments}

OW and ICC have stock in Evariste technologies. 
AT and JAW have no competing interests.

\nolinenumbers

\bibliographystyle{abbrv}

\end{document}